\documentclass[twocolumn]{aastex631}

\usepackage{float}
\usepackage{textgreek}
\usepackage{graphicx}
\usepackage{makecell}
\usepackage{lipsum}

\shorttitle{settling in low gravity}

\graphicspath{{./}{figures/}}
\usepackage{silence}

\begin{document}

\title{Experiments on Settling of Granular and Cohesive Material in Low Gravity}

\author[0000-0002-5841-2636]{Matthias Keulen}
\affiliation{University of Duisburg-Essen, Faculty of Physics,
Lotharstr. 1, 47057 Duisburg, Germany}

\author[0000-0000-0000-0000]{T. Giese}
\affiliation{University of Duisburg-Essen, Faculty of Physics,
Lotharstr. 1, 47057 Duisburg, Germany}

\author[0000-0002-7962-4961]{K. Joeris}
\affiliation{University of Duisburg-Essen, Faculty of Physics,
Lotharstr. 1, 47057 Duisburg, Germany}


\author[0000-0002-3517-1139]{J. E. Kollmer}
\affiliation{German Aerospace Center (DLR), Institute for Frontier Materials on Earth and in Space,
Linder Höhe, 51147 Köln , Germany}

\begin{abstract}

The regolith of rocky bodies, such as planets or asteroids, generally settles under gravity conditions different from those of Earth. The behavior of granular material is not easily scalable for different gravities. To predict these highly complex systems where cohesive inter particle forces can be comparable to gravitational forces, we need simulations and experiments. We did experiments on settling of three different granular samples in varying reduced gravities and examined their packing densities. We used a high precision linear stage to artificially induce reduced gravities inside the zero $g$ environment provided by the ZARM drop tower and observe the settling of our samples. The three samples were fine basalt with particle diameters of $1\text{--}200\,\mu$m, coarse basalt with $\mathbf{2\text{--}5\,}$mm and glass beads with $750\text{--}1000\,\mu$m. The artificial gravities were $150,\,250,\,500,\,750$ and $1000\,$mm/s$^2$ and therefore ranged from large asteroid gravity to almost moon gravity. We saw the granular samples have higher volumes in lower gravities and therefore lower packing densities, we also saw the fine basalt be the most sensitive to changes in gravity, up to $+19.6\,\%$ in volume for $250\,$mm/s$^2$, followed by the coarse basalt particles, up to $+12.2\,\%$ for $150\,$mm/s$^2$ and the glass beads packing density being the least sensitive to changes in gravity, up to $+4.25\,\%$ for $250\,$mm/s$^2$. With these experiments we show change in volume is not solely dependent of particle size but also roughness and uniformity, we provide real life experimental data to validate theoretical works and highlight the role of cohesive forces in low gravity environments. 

\end{abstract}



\section{Introduction} 
Considering the plans for future In-Situ Resource Utilization proposed for the return to the moon and other exploration goals one has to understand the mechanical behavior of regolith in a low gravity environment and how it differs from that on Earth \citep{Badescu.2023}.  
Predicting the mechanical packing behavior of granular matter is essential for many industrial processes here on Earth where materials across different industries such as construction, pharmaceutics or food are most commonly transported or stored in granular form \citep{Coussot.2005,Chapelle.2005b,He.2025,Kunnath.2021}. To prevent clogging, sticking or damage to the material itself for powders and grains it is crucial to know parameters such as packing density, angle of repose or cohesion \citep{Walker.1966,Raymus.1997}. \\
For future ISRU applications the dependence of these mechanical parameters of regolith and how they might differ other planetary bodies like the Moon, Mars or various rubble pile asteroids and therefore in different gravities needs to be understood \citep{Puumala.2023} \citep{Certini.2009,Zhang.2019,Mueller.2023}.\\
A reduced packing density in lower gravities has implications for many processes. Less tightly packed particles are suggested to be relevant for aeolian interaction on mars \citep{Musiolik2018} and the packing and transport of lunar regolith is of high interest for coming moon missions and the construction of an outpost \citep{Wilkinson.2005,Flahaut.2023}. The interaction with the surface of rubble pile asteroids has been shown to be highly nontrivial by missions such as Hayabusa2 \citep{Okazaki.2017} and OSIRIS-REx \citep{Beshore.2015}, which also is of great interest for the passing of Apophis in 2029 and the OSIRIS-APEX mission \citep{DellaGiustina.2023b}. \\
One of the issues with this field of study is that granular behavior is not easily scalable across different gravities. This is especially important for gravities on small bodies where we are looking at regimes where cohesive forces, for example van der Waals or electric static forces, can be comparable to that of the weight of individual particles \citep{Johnson1971,Scheeres.2010,Sanchez2014,Kimura2015,Brisset.2022}. \\
There are several theoretical works dedicated to these problems, simulating the parameters packing density, cohesion and angle of repose \citep{Elekes.2021,Persson.2022,Persson.2024} as mentioned before or more dynamical processes such as flow behavior for example in hoppers or while drilling \citep{Ozaki.2023,Fu.2026,Gaida.2025,Madden.2025}.\\
To complement these theoretical works with experimental data, we have devised a setup to investigate settling and packing in reduced gravity. For the experiments, we use three different granular samples: fine basalt, coarse basalt, and spherical glass beads, and let them settle under gravitational environments in the range of $0.1\text{--}0.01\,g$ (where $g$ is Earth's gravity). This already puts our results in gravitational regimes below what is found on the surface of the Moon.\\
We provide measurements in high quality artificial gravity for regimes not investigated experimentally before.


\section{Experiment} 
\begin{figure}[ht]
    \centering
\includegraphics[width=0.45\textwidth]{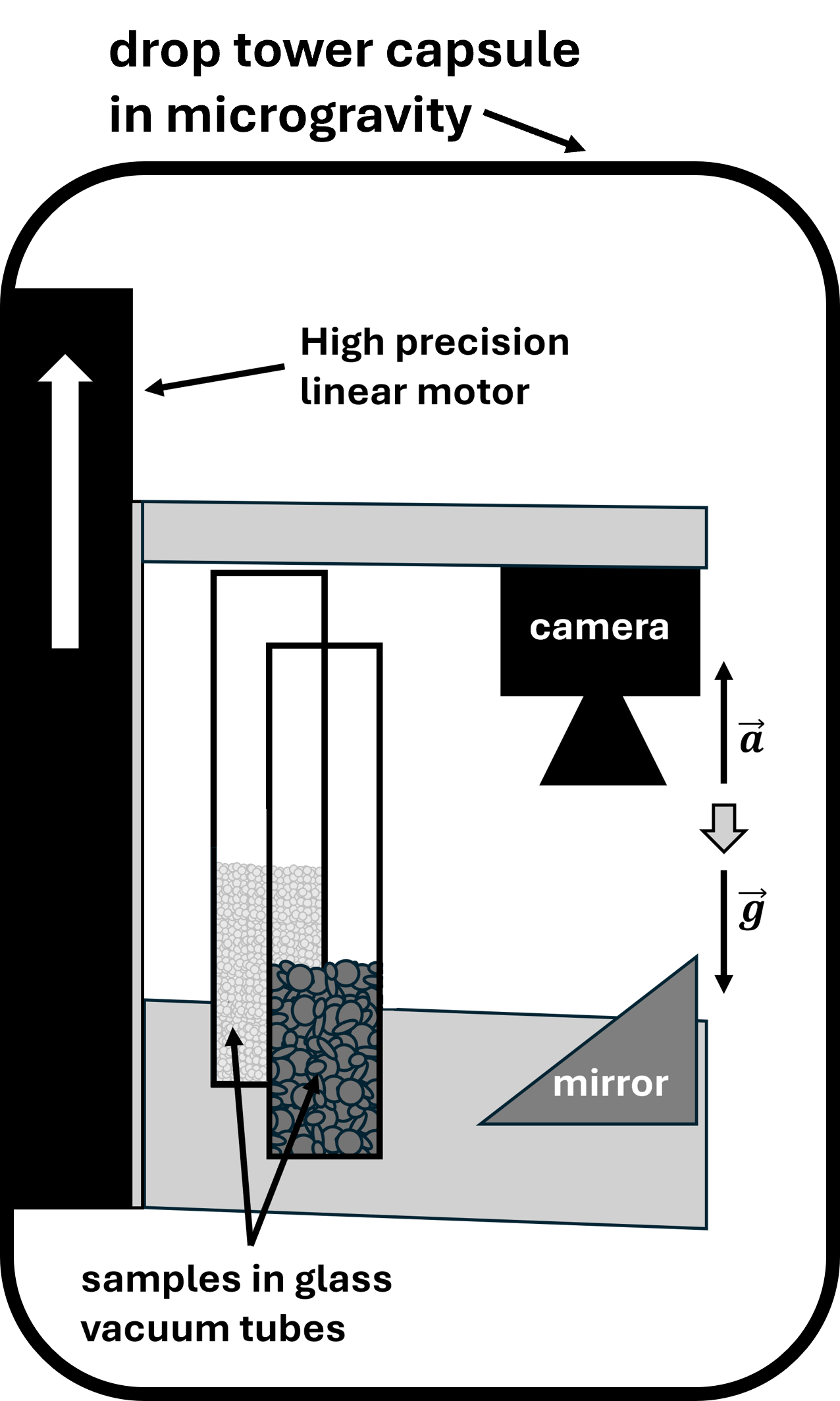}
      \caption{Set-up as used inside the Bremen drop tower capsule and the residing microgravity. With the high precision linear motor accelerating the experiment according to figure \ref{fig:traj} and with the sample tubes in view of the camera.}
      \label{fig:setup}
\end{figure}
We conducted our experiments at the drop tower facilities of the ZARM (Zentrum für angewandte Raumfahrttechnologie und Mikrogravitation) in Bremen. Here we used the GraviTower Pro Bremen (GTB), a secondary drop tower providing up to $2.5\,$s of microgravity, with residual gravities of $10^{-6}\,$g, at a time. The GTB uses a pulley system and thus eliminates the need to evacuate a vacuum chamber between flights \citep{Gierse.2017,Selig.2010,Kampen.2006}. \\
Inside the capsule provided by the ZARM team we used the setup shown in the works of \cite{Joeris.2025}, utilizing a high precision linear motor as stage inside our zero $g$ environment. By this we artificially generate the desired levels of gravity inside our experimental setup. \\
Using the ZARM drop tower ensures virtually g-jitter free zero g environments to use the linear motor in and create precise partial gravity conditions without Coriolis forces, but with high adjustability and repeatability.\\
To allow the granular samples to settle into a packing consistent with the partial gravity level the sample is vigorously shaken once partial gravity is achieved and then allowed to settle.\\
Similar work has been done by \cite{Schrapler.2015}, for lunar gravity and above, inside the microgravity environment of a parabolic flight. The parabolic flight enabled them to gather a substantial amount of data, but there is considerable residual acceleration during such flights. This poses a problem when going to even lower gravity levels and might already influence the granular sample in the gravities examined by \cite{Schrapler.2015}. Our drop tower experiments on the other hand are more time restricted, but provide very clean artificial gravities \citep{Joeris.2025}.

\subsection{Artificial Partial Gravity}
\begin{figure}[ht]
    \centering
\includegraphics[width=0.45\textwidth]{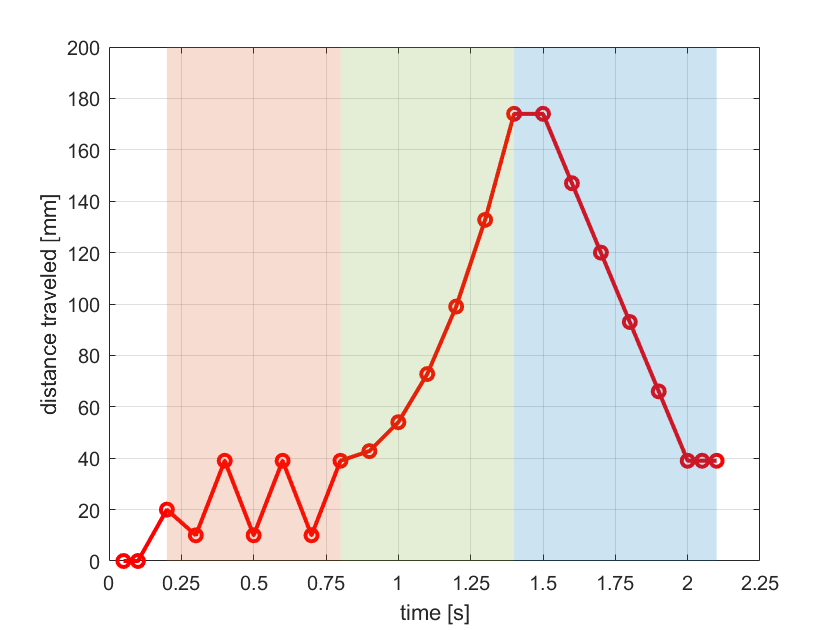}
      \caption{Example trajectory of the stage in zero $g$ for a run in $750\,$mm/s$^2$, including initializing, shake up phase (red background), partial gravity phase (green background) and the way back to the home position (blue background) to reduce the distance the stage can travel in the deceleration of the drop tower capsule. Please note that this distance traveled by the stage is completely independent from the traveling height of the drop tower capsule at any given moment. Through the acceleration of this trajectory, we reintroduce the artificial gravity back into the system inside microgravity.}
      \label{fig:traj}
\end{figure}

The setup in zero $g$ is schematically shown in figure \ref{fig:setup} with two vacuum tubes containing the granular samples and a camera pointed onto the samples using a mirror. This was necessary to minimize levers, as the drop tower capsule is decelerated at about $5\,$g \citep{ZARMFABmbH.}. All this was mounted onto the high precision linear motor (Newport IMS-LM300) or stage in our set-up. \\ 
Figure \ref{fig:traj} shows an example trajectory for artificial gravity of $750\,$mm/s$^2$ (about  $0.076g_{Earth}$). This and the other trajectories start once the drop tower capsule is in zero $g$. At first, the stage does an initialization phase followed by a shake-up before the actual artificial gravity starts. At the end the stage returns slightly above its initial position to reduce the run-up to its limit stop for the deceleration phase of the drop tower capsule. \\

We conducted $23$ successful flights in total with two evacuated sample containers observed each time. The pressure inside the sample containers ranged between $0.21$ and $0.78\,$mbar to minimize gas drag. Due to the highly time restricted nature of the experiment, not all flights resulted in fully settled samples to analyze. For this reason there is no data available for fine basalt and glass beads in $150\,$ mm/s$^2$. At this lowest gravity examined only the coarse basalt sample managed to fully settle inside the vacuum tube.\\
The experiments were divided between artificial gravities of $150$, $250$, $500$, $750$ and $1000\,$ mm/s$^2$.\\

\subsection{Samples}
\label{sec:sample}
\begin{figure}[ht]
    \centering
\includegraphics[width=0.45\textwidth]{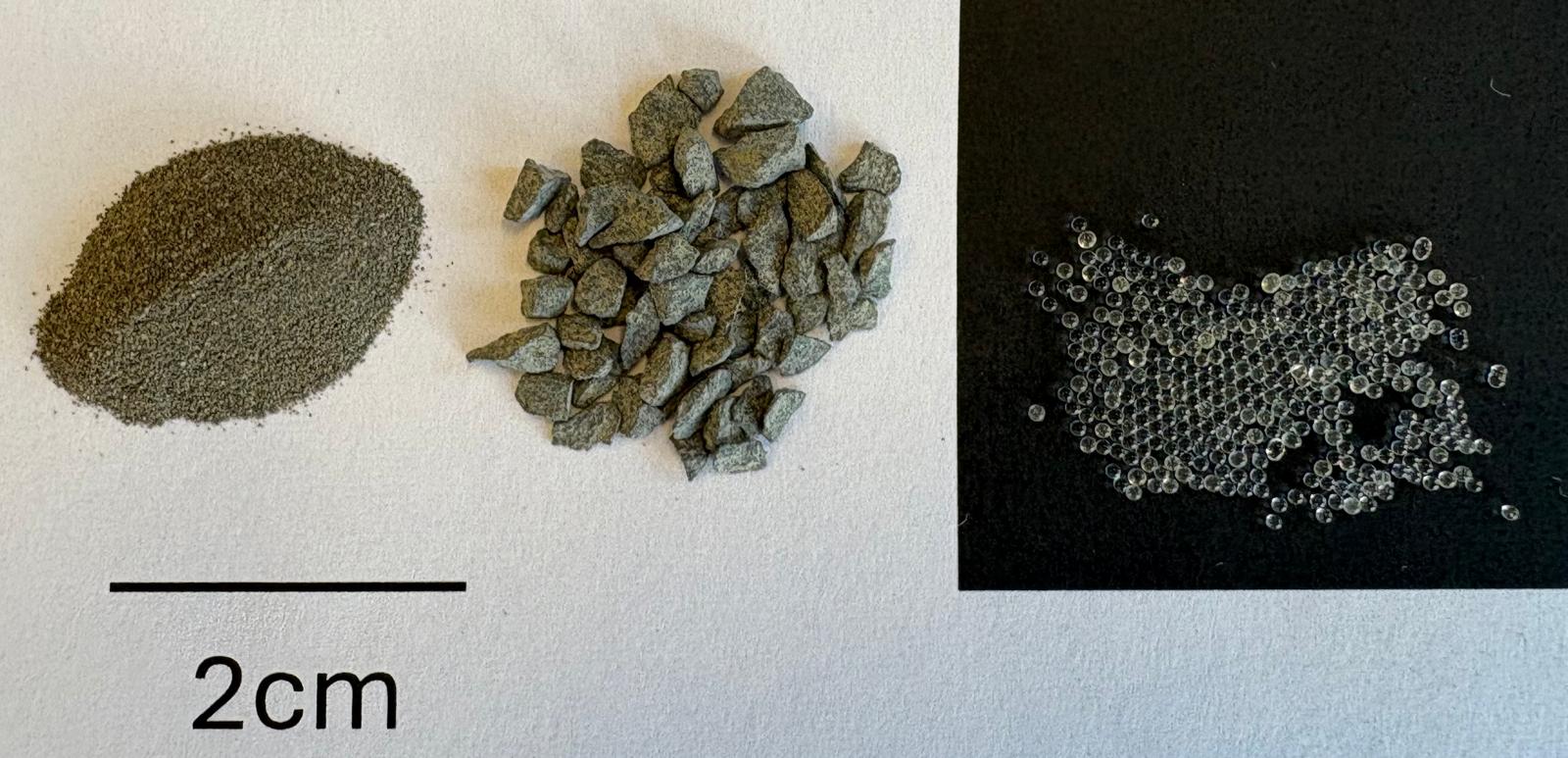}
      \caption{The different granular samples next to each other, left to right: fine basalt, coarse basalt and glass beads. Please note that the glass beads are actually lying next to the other samples on a black patch of the piece of paper lying underneath, they are not cropped in.}
      \label{fig:samples}
\end{figure}

\begin{figure}[ht]
    \centering
\includegraphics[width=0.45\textwidth]{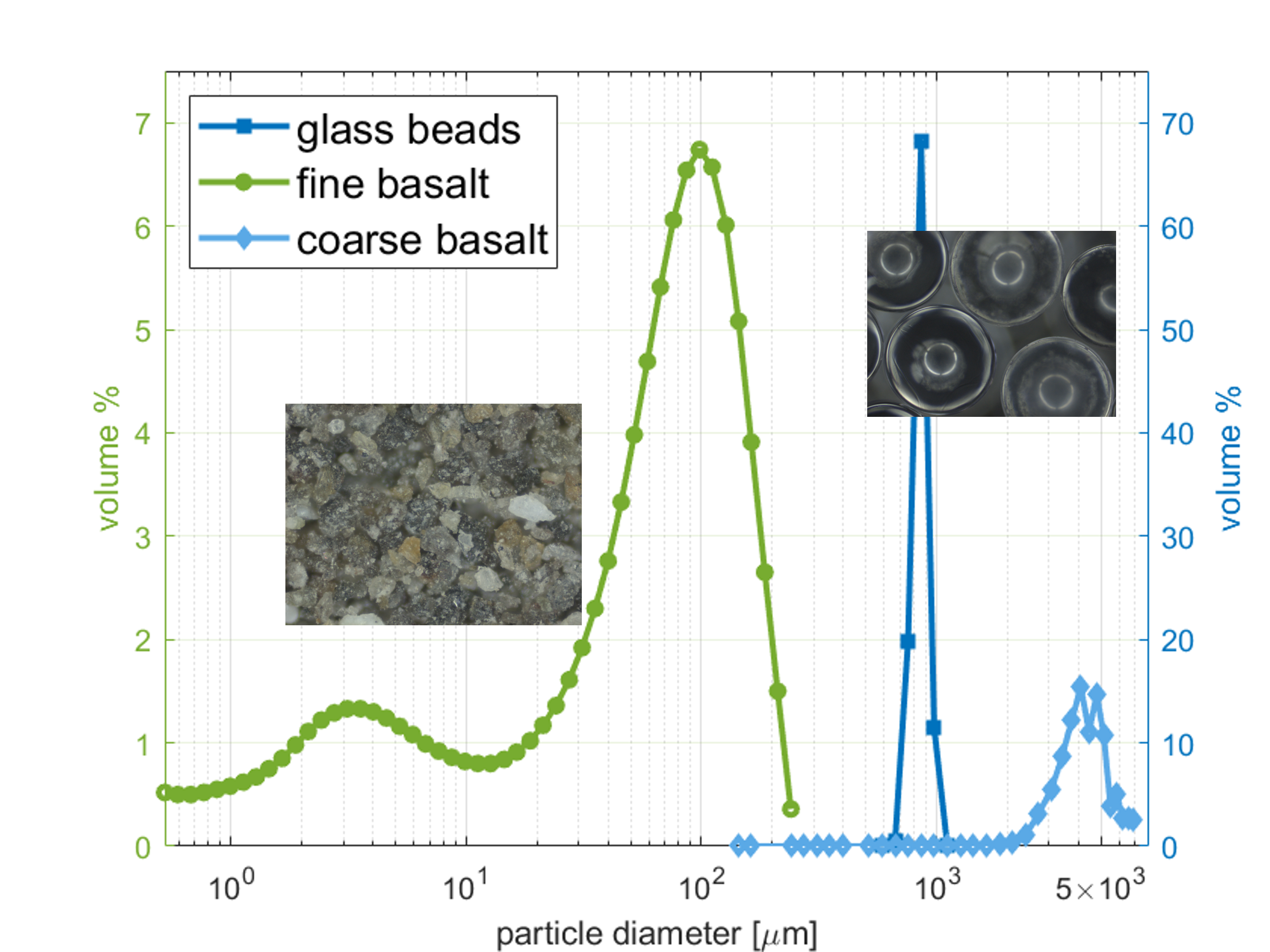}
      \caption{Size distributions of the different samples with different color coded y-axes as the Mastersizer 3000+  used here has a smaller resolution for fine grain sizes and the volume percent spreads to more measurements. Additionally insets of microscope images of the fine basalt and the glass beads are shown for roughness comparison.}
      \label{fig:sizes}
\end{figure}

We observed three different samples in these gravities, coarse basalt with diameters between $\mathbf{2\text{--}5\,}$mm and fine basalt with diameters of $1\text{--}200\,\mu$m, both bought at hardware stores and glass beads of $750\text{--}1000\,\mu$m diameter, manufactured by Whitehouse Scientific \citep{WhitehouseScientific_2005_SodaLimeGlass}, as shown in figure \ref{fig:samples}. \\
Figure \ref{fig:sizes} shows the size distributions of the fine basalt and the glass beads samples determined using a Mastersizer $3000+$. Included are insets of microscope images of the samples for roughness comparison.  \\
The coarse basalt particles were too big to be analyzed using the Mastersizer. Here we used a series of ten pictures of the sample material on a white backgroun and analyzed them using the image processing program ImageJ to determine the Feret diameter.\\
To minimize moisture and surface layers of water we dried the samples for $24\,$h at $120\,^\circ$C beforehand. We also kept the ambient pressure inside the sample containers below $1\,$mbar as long as possible. \\ 

\begin{figure}[ht]
    \centering
    \includegraphics[height=6cm]{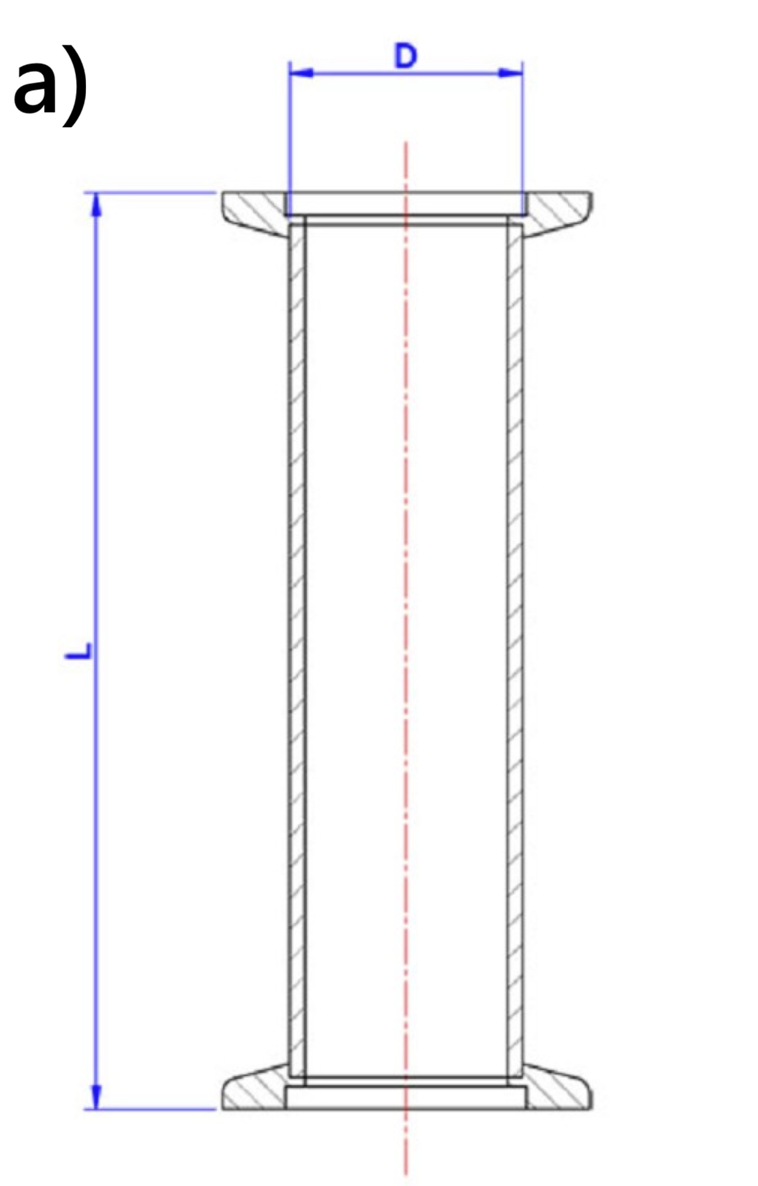}\hfill
    \includegraphics[height=6cm]{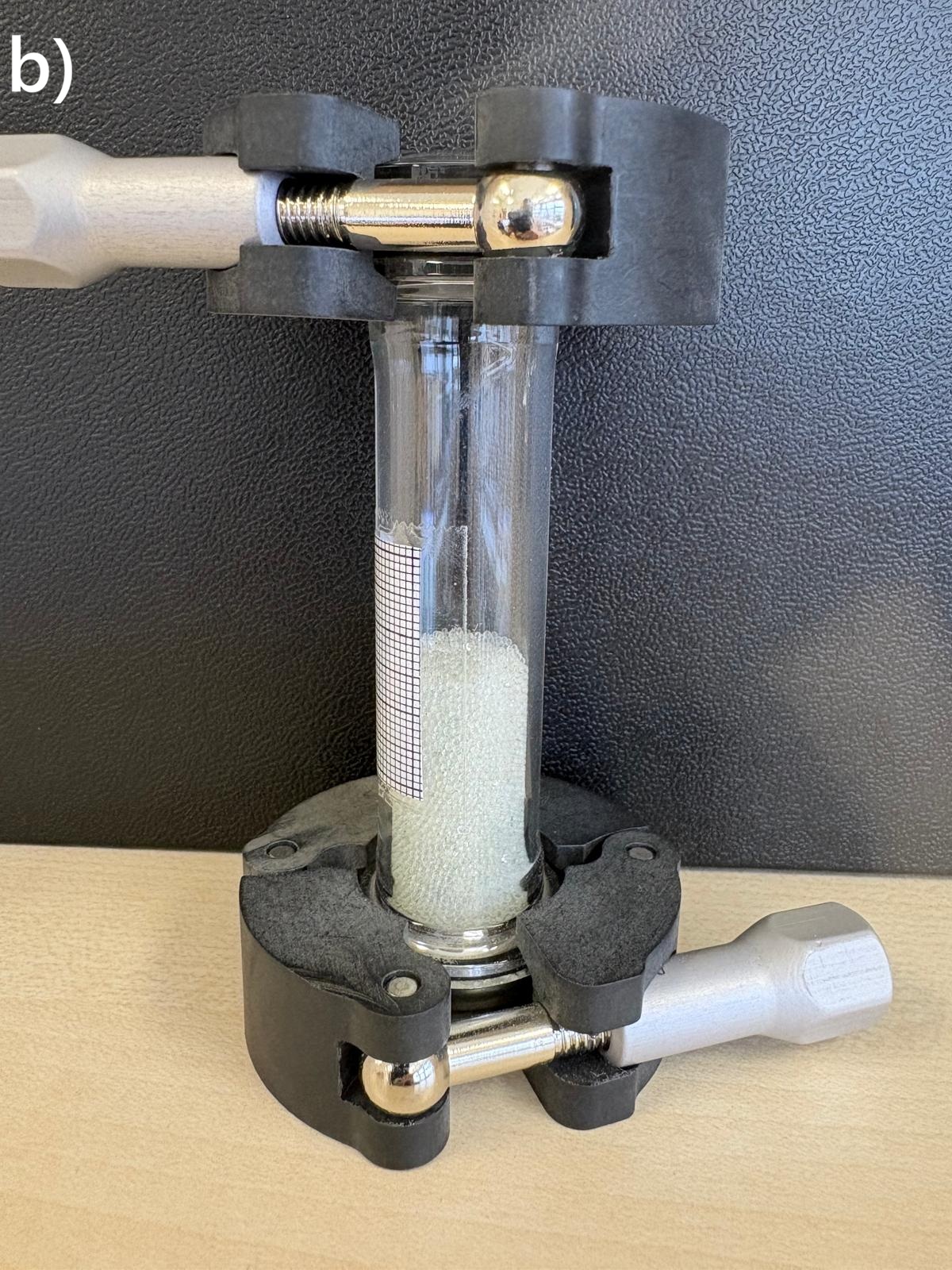}
    \caption{Sample container (a) schematic as shown on the Hositrad webshop, with total length $L$ and outer diameter $D$ \citep{hositrad_KS16GP} and (b) as used in the experiments, here filled with the $825\text{--}1000\,\mu$m glass beads.}
    \label{fig:tube}
\end{figure}

\noindent
As sample containers we used cylindrical duran glass vacuum tubes from Vacuum Technology Hositrad \citep{hositrad_KS16GP}, shown in figures \ref{fig:tube} (a) and (b), with the total length being $L=80\,$mm, outer diameter $D=20\,$mm and inner diameter $d_i=16.3\,$mm. \\

\subsection{Image Processing}
\begin{figure}[ht]
    \centering
    \includegraphics[width=0.21\textwidth]{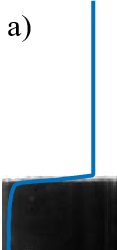}\hfill
    \includegraphics[width=0.21\textwidth]{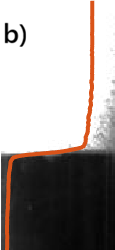}
    \caption{Comparison in filling height of fine granular basalt: 
    (a) in $1\,$g $\approx 10\cdot 10^3\,$mm/s$^2$, 
    (b) fully settled in $750\,$mm/s$^2$. Both images have the normalized brightness of pixel rows plotted in as shown in Figure~\ref{fig:pixrow}. Please note that these graphs are not the boarders of an object recognition software, they show the brightness of the pixel row, so the farther to the right the graph is, the brighter is that pixel row.}
    \label{fig:comp_fine}
\end{figure}

\begin{figure}[ht]
    \centering
\includegraphics[width=0.45\textwidth]{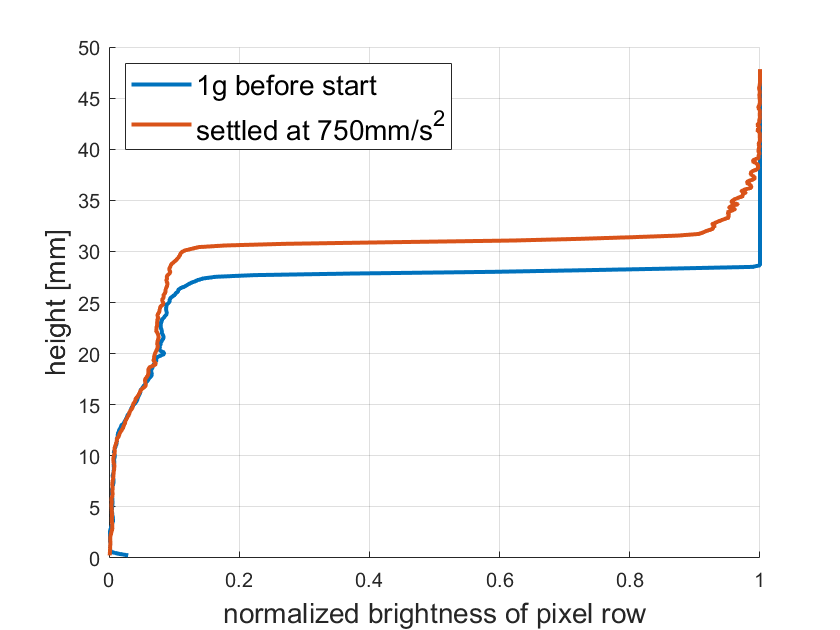}
      \caption{Normalized brightness of pixel rows of figure \ref{fig:comp_fine} (a) and (b) in comparison. Note that the y-axis is the actual pixel row converted to height in mm }
      \label{fig:pixrow}
\end{figure}

The videos obtained through these experiments were analyzed regarding the filling height of the granular samples to conclude the change in volume and therefore the packing density in $1\,$g and during the partial gravity phase. \\
Figures \ref{fig:comp_fine} (a) and (b) show the fine granular basalt sample in (a) $1\,$g before the flight and (b) fully settled in $750\,$mm/s$^2$ of artificial gravity. \\
To determine the filling height we used a matlab script to analyze the normalized brightness of every pixel row, also shown in figure \ref{fig:comp_fine} as overlay and as stand alone in figure \ref{fig:pixrow}. Here we defined the filling height at a threshold of the normalized brightness, for the basalt samples this threshold was $0.5$ of the normalized brightness, for the much brighter glass beads sample the threshold was at $0.25$. \\
Using this method we were able to determine changes in filling height $\Delta h$ converting pixels to millimeters and following that  determine changes in volume $\Delta V$ using the inner diameter of the glass tubes $d_i=16.3\,$mm with 
\begin{equation}
    V_{mg} = V_{Earth} + \Delta V =  V_{Earth} + \pi \frac{d_i^2}{4} \Delta h 
\end{equation}

with $V_{mg}$ as the total volume in partial gravity and $V_{Earth}$ the volume in Earth's gravity. While $\Delta V$ was determined in partial gravity in the sample containers through filling height, the volume in Earth's gravity $ V_{Earth}$ was calibrated in regular centrifuge tubes. This was necessary as the bottom of the sample containers with its blind flange slightly differs from a perfect cylinder. 
\\
Further we calculated the packing density $\phi$ for the different samples and gravities using the corresponding densities $\rho$ of basalt and soda lime glass and the weight $m$ of the samples, with: 
\begin{equation}
    \phi_{i,mg} = \frac{m_i \cdot \rho_i}{V_{mg}}
\end{equation}
The index $i$ refers to the different samples. For the density of basalt we used $\rho_b=3\cdot 10^{-3}\,$g/mm$^3$ \citep{ChemieDE_Gesteinsdichte_2026} and for soda lime glass $\rho_g=2.46\cdot 10^{-3}\,$g/mm$^3$ \citep{WhitehouseScientific_2005_SodaLimeGlass} and the total weights of the samples were $13.45\,$g for the coarse basalt, $13.57\,$g for the fine basalt and $14.28\,$g for the glass beads, measured using a Kern EW 820-2nM high-precision scale. 

\section{Results}
\begin{figure}[ht]
    \centering
\includegraphics[width=0.45\textwidth]{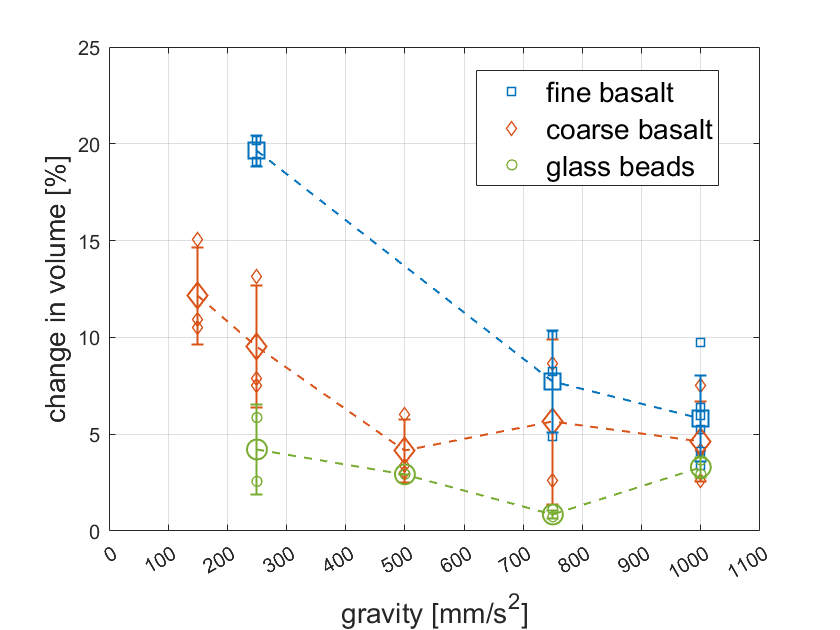}
      \caption{Change in volume in percent for different gravities. Symbols with error bars mark the averages for $150,\,250,\,500,\,750$ and $1000\,$mm/s$^2$ and the error bars are the standard deviations of these averages.}
      \label{fig:volume_perc}
\end{figure}

Using these $\Delta V$ and $V_{Earth}$ we first plotted the change in volume in percent for the different gravities and different samples as seen in figure \ref{fig:volume_perc}. 

We see an increase in volume for lower gravities as cohesive forces become more and more important compared to gravity, in other words for higher granular Bond numbers \citep{Castellanos.2005}. It is also apparent that the fine basalt sample is the most sensitive to a change in volume, showing a maximal increase in volume of to $19.6\,\%$ on average for $250\,$mm/s$^2$ of gravity. More so than the coarse basalt sample, which showed a maximal increase of $12.2\,\%$ on average but for $150\,$mm/s$^2$. \\
Notable is that the glass beads, which are closer in diameter to the fine basalt than the coarse sample, are even less sensitive to the change in gravity and only increase their volume by $4.25\,\%$ on average for $250\,$mm/s$^2$ again.\\
\begin{figure}[ht]
    \centering
\includegraphics[width=0.45\textwidth]{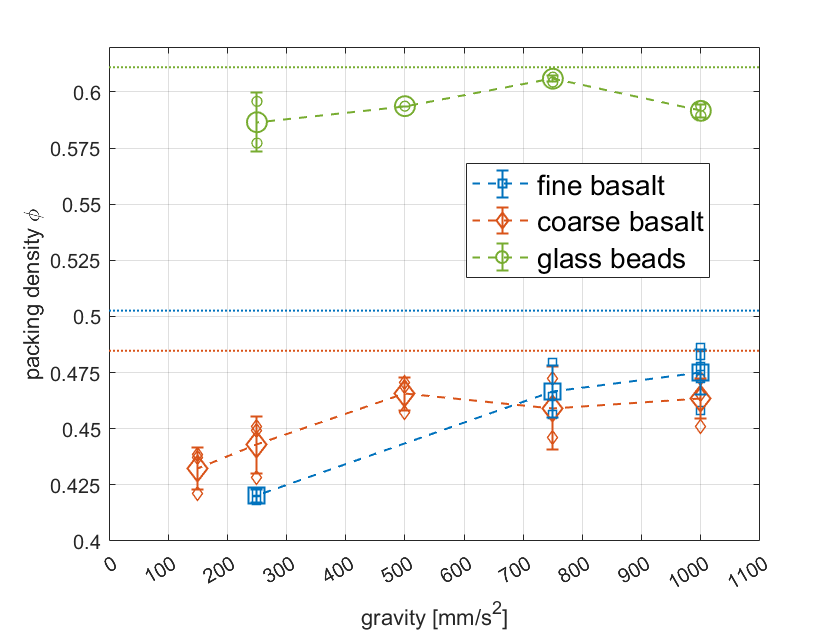}
      \caption{Packing density for different gravities. Symbols with error bars mark the averages for $150,\,250,\,500,\,750$ and $1000\,$mm/s$^2$ and the error bars are the standard deviations of these averages. The horizontal lines show values in $1\,$g.}
      \label{fig:filling_factor}
\end{figure}

In figure \ref{fig:filling_factor} we visualize the same data as packing density against the different gravities, with the additional horizontal lines showing the packing densities seen in Earth's gravity. This depiction overall does not look as impressive as the change of volume in percent, especially with all three sets of data in one picture thus broadening the range.\\ 
What it shows is the packing densities approaching the Earth's gravity values and the switch between fine and coarse basalt, with coarse basalt having the lower packing density for gravities lower than $750\,$mm/s$^2$. Therefore showing packing density is not linearly scalable for lower gravities for the same material and that particle size and roughness do play a role. \\
\begin{figure}[ht]
    \centering
\includegraphics[width=0.45\textwidth]{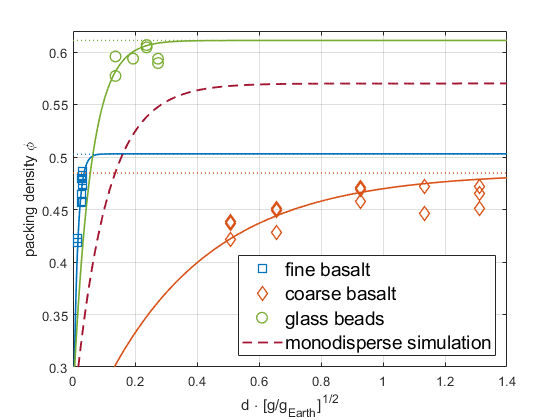}
      \caption{Packing density $\phi$ against $d \cdot \sqrt{g/g_{Earth}}$ and fitted to $\phi =  \phi_{\infty} \cdot \left\{ 1-C\cdot \exp\left[ -D_\phi^{-1} \cdot d\sqrt{g/g_{Earth}} \right] \right\}$ corresponding to the works of \cite{Elekes.2021}. Our fits are shown in the same color as individual measurements. The results of \cite{Elekes.2021} for simulated monodisperse glass particles are shown in dashed and purple. Additionally packing densities for Earth conditions are included as dotted horizontal lines.}
      \label{fig:filling_factor_scale}
\end{figure}

In figure \ref{fig:filling_factor_scale}, we fitted our data to the model proposed by \cite{Elekes.2021}, visualizing the converging behavior of packing density for high gravities and using $d \cdot \sqrt{g/g_{Earth}}$. With $d$ being the diameter of the corresponding particles, $g$ the artificial gravity and $g_{Earth}$ Earth's gravity, for the x-axis rather than just the artificial gravity to get a more standardized view for the different particle sizes. For the particle diameter $d$ we used the diameter at which the size distributions shown in figure \ref{fig:sizes} peak. \\
Further details in regards to this model are discussed in section \ref{sec:disc}.\\


\section{Discussion}
\label{sec:disc}
We find our granular samples to occupy more space in lower gravities.\\
What we were able to investigate are differences in change of volume and therefore packing density for different particles. Most prominently we see the rough granular basalt samples to be more sensitive to changes in gravity while the more uniform glass beads do not react as intensely. \\
Other than roughness the biggest difference between the samples are the particle diameters with the fine basalt at diameters around a $100\,\mu$m, the glass beads at around $850\,\mu$m and the coarse basalt even at diameters of $2000$ to $5000\,\mu$m. \\
As mentioned, to make this aspect more comparable we later used $d \cdot \sqrt{g/g_{Earth}}$ according to the works of \cite{Elekes.2021} and fitted our data to their model for the asymptotic growth function of the packing density $\phi$: 
\begin{equation}
    \phi =  \phi_{\infty} \cdot \left\{ 1-C\cdot \exp\left[ -D_\phi^{-1} \cdot d \sqrt{g/g_{Earth}} \right] \right\}.
    \label{eq:fit}
\end{equation}

Here $\phi_{\infty}$ is the associated random loose packing fraction, $C$ is a parameter set to $0.56$ as found by \cite{Elekes.2021}, basically determining how fast the curve is falling for smaller $d \cdot \sqrt{g/g_{Earth}}$ together with $D_\phi$ the characteristic length scale in millimeters. \\
The authors used DEM in their numerical simulations introduced by \cite{Cundall.1979} and improved upon since then. For further information on the simulations we refer to the original publication as well as \cite{Parteli.2014}.\\
Table \ref{Tab:fits} compares the parameters we found fitting our experimental data to equation \ref{eq:fit}. For our fits we used Mathematica's NonlinearModelFit function, the corresponding p-values are included in table \ref{Tab:fits}. For the packing density the fits converge to $\phi_\infty$ which we set at $\phi_{1g}$, the packing densities measured in Earth's gravity. Setting $\phi_\infty = \phi_{1g}$ is a good first approximation, as the simulations of \cite{Elekes.2021} suggest that packing densities do not increase much further after $1\,$g for particle sizes of our scale.
\begin{table}[ht]
\centering
\begin{tabular*}{\linewidth}{@{\extracolsep{\fill}}ccccc}
\hline
Sample/Fit & $d$ in $\mu$m & $\phi_{1g} =\phi_\infty$ & $D_\phi$ & P-Value  \\
\hline
coarse basalt & 4102.5 & 0.485 & 0.347 & 3.6$\cdot 10^{-11}$ \\
fine basalt & 98.1 & 0.503 & 0.013 & 9.8$\cdot 10^{-12}$ \\
glass beads & 859.0 & 0.611 & 0.056 & 3.7$\cdot 10^{-6}$ \\
Elekes \& Partelli & - & 0.57 & 0.103  & - \\
\hline
\end{tabular*}
\caption{Comparison between fit parameters for our findings and the theoretical works of \citep{Elekes.2021}. With $d$ being the particle diameter where the size distribution peaks in $\mu$m and $\phi_{1g}$ the measured packing density of the samples in Earth's gravity.}
\label{Tab:fits}
\end{table}

Resulting in $D_\phi$ being our last free fit parameter, containing information about particle interactions and cohesion. \\
This model does, however, not include a size distribution of a significant width. The simulations of \cite{Elekes.2021} were done with monodisperse spherical particles. Although our glass beads are about as monodisperse as we can get in experiments, the size distribution still shows a significant spread around $850\,\mu$m, as seen in figure \ref{fig:sizes}, just like the fine basalt particles.  the size distribution of the coarse basalt spreads even between $2$ and $5\,$mm. For this reason our data points do not collapse onto one curve as in the theoretical studies of \cite{Elekes.2021}.\\
As mentioned \cite{Schrapler.2015} did similar work focusing on gravity values between $1\text{--}0.1\,g$ and the dependency of filling fraction on sample depth. They examined three samples, glass spheres of $60\,\mu$m diameter, spherical soda lime particles of $1\,$mm diameter and agglomerates consisting of $1.5\,\mu$m SiO$_2$ spheres. Their experimental data for gravities just above those examined in this study does not show a clear trend for the dependence  of granular filling factor on gravity. The reason for this might be, that they are outside of the regime of noticeable change. When looking at our fits in figure \ref{fig:filling_factor_scale} for the fine basalt and the glass beads, which are the most similar samples to theirs, an increase in gravity would bring the filling factor close to the associated random loose packing fraction $\phi_\infty$. \\
The residual acceleration of the aircraft in parabolic flight experiments might be another reason. This residual acceleration of parabolic flights typically is in the range of $\pm 0.05 g$ as absolute values \citep{Pletser2016}, leading to a significant error, when dealing with $0.1 g$.  It might also explain their generally higher values for filling fractions of $0.5\text{--}0.7\,g$ for the samples other than the agglomerates. The g-jitter could work like a tapping on the granular samples, compressing them further after the settling in low gravity \citep{Arsenovic2006}. \\
In situ measurements of lunar regolith reported by \cite{10115961} show a porosity of $>40\,\%$ for regolith depths of $<10\,$cm, corresponding to packing fractions of $<0.6$. Incomparison,  our measured packing densities $\phi$ are around $0.425$ to $0.475$ for the basalt samples and around $0.6$ for the glass beads. This generally agrees with our findings considering that our artificial gravity levels only go up to about $2/3$ of lunar gravity.\\
In situ measurements for regolith packing density on asteroids are hard to come by, as most are either bulk densities of whole bodies \citep{Grott2020Ryugu} or thermal measurements of only the optically active layer \citep{VERNAZZA20121162}. \cite{Magri2001} estimate a mean near surface porosity of several asteroids of $51\pm 14 \,\%$, corresponding to packing densities of $0.49\pm0.14$. This again is in general agreement with our findings, even though the considerable spread of values between different asteroids has to be mentioned. They estimate a porosity for example for Ceres of $67\pm4\,\%$ and for Eros of $47^{+26}_{-16}\,\%$. The reported gravities of these bodies range from $250\,$mm/s$^2$ for Ceres to $2.3\,$mm/s$^2\text{--}5.6\,$mm/s$^2$ for Eros \citep{Thomas2005Ceres,MILLER20023}.\\
We note that humidity and surface layers of water can play a big role in cohesion and overall handling of granular material, as shown by \cite{Persson.2024}, we tried minimizing these effects by drying the samples as reported before in section \ref{sec:sample}. 
Another component of cohesion can be electrostatic forces through tribocharging. Here, even single impacts can generate significant charges \citep{Keulen.2025,Becker.2024a,Grosjean2023} , while whole granular systems have been shown to also charge \citep{Waitukaitis2014,Lacks2011,Carter.2019} and to be able overcome growth barriers in protoplanetary discs by shaking and charging \citep{Teiser.2025,Teiser.2021}. Signs of that can be seen in figure \ref{fig:comp_fine} (b) with some particles sticking to the glass container. This impairment is virtually impossible to fully remove in experimental setups, as tribocharging is inevitable in shaken granular samples \citep{Lacks2019,CARTER2020,Jungmann2022x}. \\
Still, trying to reduce tribocharging might give additional insight on the role of static electricity and is one opportunity to build upon this work. Feasible options for this include coating the sample and the container in conducting material or using ionized gas to neutralize the particles \citep{Sarkar.2012,LaMarche.2009,Choi.2013,Haeberle2018,Shi.2025}. As these methods need extensive testing, this requires a dedicated series of microgravity experiments and is an ineresting research question of its own, but beyond the scope of this work.\\
It should also be noted that the sample containers were in part chosen because of limited space and boundary effects might play a role. 
For example, if we compare our set-up with granular flows through a hopper, a generally more broadly studied problem, critical orifice sizes are typically $5\text{--}10\,$times the particle size \citep{Zuriguel.2020,Drescher.1995,Jenike.1964}.\\
If we compare our particle diameters $d_p$ with the inner diameter $d_i$ of the sample container we get ratios $d_i/d_p$ of the samples of about $\mathbf{4.98}$ for the coarse basalt, $\mathbf{166.16}$ for the fine basalt and $\mathbf{18.98}$ for the glass beads. Therefore, the coarse basalt might be influenced by boundary effects, although a significant difference from hopper flow is that we do not have a downward narrowing of the container. 

\section{Conclusion}
We have observed the change in volume and therefore in packing density in reduced gravities from $150$ to $1000\,$mm/s$^2$ of three different granular samples, fine basalt with particle diameters of around $1\text{--}200\,\mu$m, coarse basalt at $\mathbf{2\text{--}5\,}$mm and glass beads with diameters of $750\text{--}1000\,\mu$m. \\
We did so in controlled partial gravities utilizing a high precision linear stage and the zero gravity environment of the ZARM  GTB Pro drop tower Bremen. Through this we greatly decreased noise and accelerations by jitter and altogether avoided Coriolis forces common in other setups for reduced gravity, such as centrifuges on parabolic flights. \\
We see  increased sensitivity to lower gravities for rougher particles and for finer granular material, highlighting the importance of cohesive forces for real world applications.\\
With this we provide experimental data to validate theoretical works and simulations. 

\begin{acknowledgments}
This project is supported by DLR Space Administration with funds provided by the Federal Ministry of Economic Affairs and Climate Action (BMWK) under grant number 50WM2243. We acknowledge helpful discussions with Eric Parteli and Jens Teiser. 
\end{acknowledgments}

\bibliography{bib}{}
\bibliographystyle{aasjournal}

\end{document}